# High-Efficiency Photodetector Based On CVD-Grown $WS_2$ Monolayer

Rakesh K. Prasad[1], Koushik Ghosh[2], P. K. Giri[2], Dai-Sik Kim[3], Dilip K. Singh[1*]

[1]Department of Physics, Birla Institute of Technology Mesra, Ranchi -835215, India

[2]Department of Physics, Indian Institute of Technology Guwahati, Assam-781039, India

[3]Department of Physics and Quantum Photonics Institute and Center for Atom Scale Electromagnetism, Ulsan National Institute of Science and Technology (UNIST), Ulsan-44919, Republic of Korea

[*]Email: dilipsinghnano1@gmail.com

Abstract

Future generation technologies demand high efficiency photodetectors to enable sensing and switching devices for ultrafast communication and machine vision. This require direct-band gap materials with high photosensitivity, high detectivity and high quantum efficiency. Monolayered two- Dimensional (2D)-Semiconductors based photodetectors are the most promising materials for such applications, although experimental realization has been limited due to unavailability of high quality sample. In the current manuscript, we report about $WS_2$ based photodetector having sensitivity of 290 $AW^{-1}$ upon 405 nm excitation and incident power density as low as 0.06 $mW/cm^2$. The fabricated device shows detectivity of $52 \times 10^{14}$ with external quantum efficiency of $89 \times 10^3$ %. The observed superior photo-response parameters of CVD grown $WS_2$ based photodetector as compared to Si-detectors establishes it capability to replace the Si-photodetectors with monolayered ultrathin device having superior performance parameters.

**KEYWORDS:** Monolayer Tungsten disulfide, 2-Dimensional growth, CVD, High efficiency Photodetectors,

Photodetectors (PDs) are important component to achieve devices with multi-functionality which convert photons into electrical signals. They are widely used for imaging[1], optical communications[2], sensing[3] and in biomedical instruments[4]. There have been overgrowing demand to achieve PDs with enhanced performance in terms of high sensitivity over wide spectral range, faster switching rates and high pixel density requiring small device foot prints. These requirements are further intensified with applications requiring PDs operation under different environments like high temperature, low light conditions requiring extremely high signal/ noise ratio, thermal imaging using PDs for night vision. Meeting these desired properties requires replacement of conventional materials like silicon / Germanium with new materials having direct-band gap structure and strong light matter interaction.

Atomically thin layered materials like graphene and Transition metal dichalcogenides (TMDCs) due to their high photo absorbance properties are considered as one of the potential candidates to achieve efficient PDs[5,6]. A number of 2D- materials, their hetro-structures and numerous device architectures has been experimentally explored to meet the desired properties [7,8,9,10]. PDs based on 2D materials have broad electromagnetic spectrum sensitivity as compared to PDs based on traditional bulk semiconductor due to wide bandgap range 0 to 6 eV[11]. Among wide range of 2D materials, transition metal dichalcogenides (TMDCs) has been explored extensively due to their decent mobility, direct-band gap structure resulting from quantum confinement[12,13], high quantum efficiency [14] and the possibility of fabrication of hetro-structures without requirement of epitaxial growth[6,15]. 2D-TMDs are formed by (Mo, W) transition metals atoms and chalcogenides (S, Se, Te) forming $MX_2$ semiconductors with a bandgap in the range 1.1ï 2.1 eV (visible to infrared) region[6]. The in-plane X-M-X covalent bonds lead to the formation of isolated atomic layers. The individual X-M-X sheets engage with one another via weak van der Waals forces resulting in bulk crystals[16,17,18]. In the monolayer limit, these materials apart from showing direct band-gap electronic structure show strong excitonic effects and spin and valleyï dependent properties making them highly desirable for photonics and optoelectronics applications[19].

Among the TMDs family of materials, Molybdenum disulfide ($MoS_2$) and Tungsten disulfide ($WS_2$) have been most extensively researched for photodetector applications[19]. The first monolayered $MoS_2$-based phototransistor showed good stability and a response time of 50 ms due to low absorbance[20]. $WS_2$ shows phonon-limited high electron mobility of 1,103 $cm^2$/ (V.s) at room temperature[21], broader spectral response as compared to GaAs and Si (less than 1 nm thickness) [22,23]. It has high optical absorption coefficient of $10^5$-$10^6$ $cm^{-1}$ [22]. It shows layerï dependent band gap of 1.2 ï 2.0 eV [23] and large exciton binding energies of 700- 800 meV[14,23]. Monolayered $WS_2$ shows photoluminescence quantum yield as high as 2 %, which is much higher than that of suspended monolayer $MoS_2$ 0.42% [24]. These properties makes $WS_2$ preferred material for photodetector applications as compared to other 2D TMDs. In one of the early attempts, photodetector fabricated using mechanically exfoliated $WS_2$-nanosheets exhibited high responsivity in the range of 0.1 A/W to 5.7 A/W over the spectral range of 370 to 1064 nm[25,26,27]. Due to the limited lateral dimensions, the devices fabricated using mechanically exfoliated flakes (~100 micrometers) having random shape, thickness and crystal quality shows wide variation in the device properties [26] .This limits its use towards imaging pixels and display devices, which normally require large arrays of photodetectors.

Recently, there has been an ongoing effort to grow continuous large crystallite $WS_2$ using the chemical vapor deposition (CVD) technique [24,28,29], although only a few have been able to demonstrate successful growth of continuous mono-layered $WS_2$ having quality suitable for the fabrication of optoelectronic devices[30]. Unfortunately, the fabricated phototransistor showed poor responsivity limited to 18.8 mA $W^{-1}$ at 532 nm by applying a high voltage (~ 60V) through the gate channel in the vacuum[31]. Any practical application of a device with such high applied potential (+ 60 V) is not feasible. In another attempt, the three terminal $WS_2$ photodetectors showed responsivity = 0.52 mA/W, while the photodetector fabricated on the flexible substrate showed responsivity of = 5 mA/W [32] with 532 nm laser excitation. Recently, in 2021 Ying Chen fabricated two terminal devices using grown monolayer $WS_2$ triangular flakes with the high photo-responsivity of 7.3 $AW^{-1}$ with 500 nm[33]. In 2016 Ying Chen et al. grew monolayer $WS_2$ via Silanization treatment and fabricated two terminal photodetector devices which give responsivity upto 307 $AW^{-1}$ under 500 nm excitation[34]. In another report, mechanically exfoliated $WS_2$ nanoflakes utilizing surface-adsorbates-sensitive charge transport and under $NH_3$ gas an external quantum efficiency of 884 $AW^{-1}$ was observed. In the case of CVD grown $WS_2$ monolayers, the observed limited photosensitivity and slow response time requires further increase in the crystallite size and continuity of the grown material.

In this work, we report about synthesis of large-area continuous films of $WS_2$ monolayer through CVD and observed superior photo-detection parameters of the fabricated photodetector based on grown $WS_2$ monolayer.

For $WS_2$ growth Tungsten trioxide $WO_3$ 20 mg (99.9 % sigma Aldrich) and sulfur powder 200 mg (99.9% Sigma Aldrich) were used as precursors and carrier gas Ar flow rate was kept at 100 sccm. The schematic of the CVD setup and the parameters of $WS_2$ growth are shown in supporting Information figure S1. The growth was carried out at 850 °C for 15 minutes using a single zone CVD furnace at atmospheric pressure.

The growth of $WS_2$ flakes was initially confirmed using an optical microscope (Leica 580) with an objective lens 10× (NA 1.3), 50× (NA 2.6). Detailed morphology of grown samples was recorded using a field emission scanning electron microscope (ZEISS Sigma 300 FESEM). Raman spectroscopy ((LabRam HR800, Jobin Yvon) in the Raman shift range 350-450 $cm^{-1}$ was used to determine the number of layers of $WS_2$. Photoluminescence spectra was recorded in the range 500-800 nm using an $Ar^+$ ion laser ($ə_{ex}$ = 532 nm). Atomic Force Microscopy (AFM) (Cypher, Oxford Instruments) in noncontact mode was used to measure the thickness and surface topology. $WS_2$ monolayer-based photodetectors were fabricated using $WS_2$ on $SiO_2$ (285 nm)/p-Si wafers. On top of grown $WS_2$, microelectrodes were fabricated using shadow masking. The gap between adjacent electrodes was maintained at 30 εm. The Iī V characteristics and temporal response of the photocurrent were determined using a microprobe station (ECOPIA EPS-500), a 405 nm laser with TTL modulation, and a source meter (Keithley 2400).

Figure1 (a) shows the optical micrograph of $WS_2$ grown using CVD. It shows continuous growth over the complete wafer (1.5 × 1.5 $cm^2$) with a coverage area of 98 % calculated using Image J software. It shows small triangular regions of nucleation centers, with dark colour contrast indicating nucleation centers for more than one layer structure. FESEM image, figure. 1(b) shows magnified image of as-grown single crystal of $WS_2$ having size of ~ 80 εm. Figure 1 (c) shows the Atomic Force Microscope (AFM) image of the $WS_2$ isolated crystal and figure 1(d) shows the thickness profile graph of the $WS_2$ monolayer is ~ 0.9 nm.

Figure 2(a) shows the Raman spectra of the $WS_2$-grown sample. Spectra were fitted, using Lorentzian line shape to ascertain the peak profile parameters. It shows peaks at 294.5 $cm^{-1}$, 323.0 $cm^{-1}$, 343.1 $cm^{-1}$, 351.0 $cm^{-1}$, 355.6 $cm^{-1}$ and 417.9 $cm^{-1}$ peaks. The peak at 343.1 $cm^{-1}$, 355.6 $cm^{-1}$ arises due to first order in-plane vibration $E^1_{2g}$(M), $E^1_{2g}$(ũ) mode [35] and 417.9 $cm^{-1}$ is first order out-of-plane vibration $A_{1g}$ (ũ) mode. Other peaks observed at 294.5 $cm^{-1}$ and 323.0 $cm^{-1}$ are attributed to the combination modes of 2LA-2$E^2_{2g}$ and 2LA-$E^2_{2g}$, respectively. The frequency separation of 62.3 $cm^{-1}$ between the $E^1_{2g}$(ũ) and $A_{1g}$(ũ) [28] and the strong 2LA(M) mode at a laser excitation of 532 nm caused by the double resonance scattering could be the spectral fingerprint of monolayer $WS_2$ .[36] The first order Raman spectra of the $WS_2$ monolayer shows two optical phonon modes at the Brillouin zone center (□□□□ (ũ) and $A_{1g}$ (ũ)) and one longitudinal acoustic mode at the M point LA(M). □□□□ (□) was an in-plane optical mode, $A_{1g}$ (ũ) correspond to out-of plane vibration of the sulfur atom as the longitudinal acoustic phonons LA(M) were in ï plane collective movement of the atom. Additional peaks correspond to the second order Raman mode which are multi phonon combination of these first-order modes. Figure 2(b) shows the photoluminescence (PL) spectra from as grown sample along with fitted peaks using Gaussian line shape. Strong PL peak between 600 and 680 nm indicates monolayer $WS_2$ crystal growth [28]**.** The peak has maxima at the wavelength of 632 nm (direct band gap 1.96 eV) which falls in the range of reported PL peak positions for monolayer $WS_2$ thin film [18,25,26]**.** Spectra shows two Gaussian components at 632.5 nm and 638.5 nm arising from A exciton (X) and Trions ($X^-$)[37,38]**.**

The schematic diagram of the fabricated device is shown in Figure 3(a). Figure 3(b) shows the photoresponse of the device under dark and various laser excitation wavelengths: 360 nm with an incident power density (4.65 $mW/cm^2$), 405 nm (52.3 $mW/cm^2$), 532 nm (80 $mW/cm^2$) and 630 nm (2.25 $mW/cm^2$). The device showed an extremely low dark current (~ 9.5 pA). Figure 3(c) shows the I-V response with varying incident photon density in the range of 0.06 mW/cm$^2$ to 75.3 mW/cm$^2$ for λ= 405 nm excitation. The variation in the photocurrent for the applied potential of +5V is shown in Figure 3(d). The photo-induced carrier concentration increases exponentially with increasing incident photon density. The induced photocurrent ($I_{ph}$) follows $I_{Ph} = 5*10^{-7} - I_o * \exp(R*P)$ dependence with $R$ = -0.821 and $I_o = 2.062*10^{-7}$. Where P indicates incident power density. The responsivity first increases fast and then saturates with increasing incident power density. Due to the complex interplay of the processes of electron-hole generation, trapping and recombination deviation from the linear dependence is observed. This is different from the previous observation where linear response was observed[25,26].

The change in the photocurrent with varying incident laser power (λ = 405 nm) at a constant bias of 5V is shown in figure 4(a). With increasing irradiation power in the range 0.06 ï 75.3 mW/cm$^2$, the photocurrent sequentially changes from 10 nA to 0.12 μA. Through rise and fall time of the photodetector and its dependence on incident laser power irradiations, the dynamic behavior of the photodetector was explored. Observed experimental data of the rise and fall times of the photocurrent (I (t)) was fitted using single exponential function given by $I(t) = I_0 + A \exp^{-t/Ű}$, Where $I_0$ and A are the constants and Űis referred as time constant. The photocurrent growth and decay shows systematic growth with the variation of laser power illumination. The rise time varies from (0.20 to 0.60 sec) and the fall time varies from (0.07 to 0.27 sec) as shown in figure 4(b). Figure 4(c) shows the temporal response of the photocurrent under varying applied bias voltage with a fixed laser power intensity. With increase of bias potential the photocurrent systematically increases. The Rise time varies from (0.18 to 0.83 sec). Interestingly with increasing bias potential upto 3V the rise time sequentially increases and upon further increase of applied bias voltage (to 4V and 5V) the rise time decreases. Similarly the fall time sequentially increases from 0.07 sec to 0.28 sec with increasing bias voltage for constant illumination power (52.4 $mW/cm^2$).

The performance of the fabricated photodetector is estimated in terms of responsivity (R), detectivity (D*) and external quantum efficiency EQE (%) **[39].** The responsivity is calculated using $R_\lambda = \frac{I_{Ph}}{P_{in} A}$. Where $I_{Ph}$ represents the photocurrent ($I_{Light}$ - $I_{Dark}$), $P_{in}$ represents the power density of incident light, A the effective active area and photosensitivity is represented in terms of ratio $I_{ph}/I_{Dark}$. Further, detectivity (D*) was calculated using D* = $\frac{\square_{\square}\sqrt{\square}}{\square(\overline{\square\square\square_{\square}\,\square\square\square})}$. The number of carriers produced per incident photon is expressed as EQE and is calculated: EQE (%) = ($1240*\frac{\square_{\square}}{\square}$)*100, Where $\square$ represents the operating wavelength**[40]**. We observed the responsivity of 290 $AW^{-1}$ at 405 nm of excitation and with incident power density of 0.06 $mW/cm^2$. The observed responsivity is much higher than previously reported values even for the PLD grown samples (0.70 $AW^{-1}$) due to their poor control of number of layers **[25].** Few of the experimental attempts have shown comparable responsivity (100-570 $mAW^{-1}$) for the bilayers or, multilayer $MoS_2$ / $WS_2$ primarily due to increased optical absorption (40-85 %) as compared to the monolayered structures (2-10 %) **[25,41]**. The variation in the responsivity (R), detectivity (D), and external quantum efficiency (%) with varying incident power density ($mW/cm^2$) and voltage (V) are shown in Figure. 5(a) and Figure 5(b) respectively. At 0.06 $mW/cm^2$ the photodetector shows responsivity (290 $AW^{-1}$), detectivity ($52\times 10^{14}$) and external quantum efficiency of $89\times 10^3$ %. For the incident power range of 0.1 $mW/cm^2$ -10 $mW/cm^2$ range responsivity, detectivity and external quantum efficiency remains nearly constant. Detectivity represents the ability to detect weak signals. For estimating detectivity, short noise is assumed to dominate the dark current. Beyond 10 $mW/cm^2$ incident power range, these figures of merits monotonically decreases with increasing incident power density. As the $P_{in}$ increases, more holes fill shallow trap states where the lifetime is short. This results in faster recombination, hence R decreases. When $P_{in}$ increases, $D^*$ also decreases and $I_{dark}$ remains unchanged. Figure 5(b) shows a monotonically increasing figure of merit with increasing forward bias potential of the device at a given incident power density (52.4 $mW/cm^2$).

The generation mechanism of photo-current can be understood in terms of (MïSïM) architecture as illustrated in figure S2(a). For electron transfer, charge carriers must overcome Schottky potential barriers created at semiconductor-metal interfaces. The choice of metal plays a significant role in the electrical contacts on the height of the Schottky barrier, here we have chosen the Aluminum metal for electrical contacts due to its work function (ű = 4.5 eV) nearly equal to the $WS_2$ work function (ű = 4.6 eV )[42]. Since the Schottky barrier acts as a current rectifier, the electron can move from semiconductor to metal and is controlled by two sides of the interface in equilibrium (contact barrier height) if there is no external effect on the system. Under dark conditions and without applying any bias voltage, the device exhibits a symmetric Schottky structure under the same vacuum level. Under illumination and without bias voltage, the barrier height depend on the generation and recombination of photo-generated charge carriers, when the photodetector devices absorb the light. The excitation of an electron or other charge carriers to a higherï energy state produces an electrical potential separation in the metal-semiconductor junction and the photocurrent can take a direction. Application of positive bias voltage changes Schottky barrier height along with the direction and magnitude of the photo-current schematically as shown in Figure S2(b). Under both bias voltage and illumination, the photo-generated charge carries start distribution at metal/ semiconductor interfaces. Depending on the charge accumulation process, the Schottky barrier height of the two interfaces tends to vary further. Photo-excitation and application of biasing potential allows the magnitude of the electric field to increase further.

The switching response of the fabricated photodetector was measured at 2 Hz of excitation frequency with 405 nm illumination (power = 0.06 $mW/cm^2$) as shown in Figure S3 (Supplementary information). We observed an on/off ratio of $10^3$ at biasing voltage of 5V upon 405 nm excitation. The observed switching time of $T_{rise}$ = 0.20 sec and $T_{fall}$= 0.07 sec is faster than the previous report by J. D. Yao et al. in year, 2015 at biasing of 1V [25]. For the PLD-grown $WS_2$, under vacuum conditions, they observed a switching time of $T_{rise}$ = 9.9 sec and $T_{fall}$= 8.7 sec. The comparison of the key photodetector performance parameters reported by various researchers for $WS_2$ based PDs and Silicon PDs (for comparison purpose) has been enlisted in Table-I (Supplementary information). Typically the photoresponse time of Si based photo-detectors ranges in the range 0.2-5.9 μs much faster than the observed photoresponse time of $WS_2$ based photodetectors (~few milli-seconds), While $WS_2$ based photodetectors, shows much higher responsivity 290-3106 $AW^{-1}$ (Our work and Yang, R.et al, *ACS applied materials & interfaces* **2017**) as compared to Si based photodetectors 0.14 $AW^{-1}$ [43]. The device exhibits internal gain because of the difference of the recombination lifetime and transit time [44]. Since the electrons travel faster than holes and the recombination lifetime is very long, during transport in photodetectors electron completes its trips sooner than the hole.

In summary, monolayered continuous $WS_2$ was grown by atmospheric pressure CVD having crystallite size of ~80 μm using single zone furnace. Photodetector based on it showed responsivity of 290 $AW^{-1}$ at incident power density of 0.06 $mW/cm^2$ and excitation wavelength of 405 nm. For the first time such high responsivity has been observed in CVD grown monolayered $WS_2$ sample. We also observed faster switching time of about $10^3$-$10^4$ μs for our CVD grown samples as compared to previous reports. Defect concentration and grain boundaries impregnated during growth of $WS_2$ limits their switching time. Observed comparable switching time of ~ $10^3$ μs and high detectivity (D = 52 $\times 10^{14}$) indicates good quality of the samples and possibility of realizing efficient photodetectors based on monolayered $WS_2$ grown by atmospheric pressure CVD.

## ACKNOWLEDGMENT

Dilip K. Singh thanks UGC-DAE CSR Indore (CRS/2021-22/01/358), NM-ICPS ISI Kolkata and DST, Government of India (CRG/2021/002179; CRG/2021/003705) for financial support. Prof. Dai-Sik Kim thanks National Research Foundation of Korea (NRF), Korea government (MIST) for the grant (NRF-2015R1A3A2031768 NRF-2022M3H4A1A04096465). We are also thankful to the CIF, BIT Mesra, INUP programme of IIT Guwahati and UNIST Central Research Facilities (UCRF) for the support of their facilities and equipment.

**Figures:**

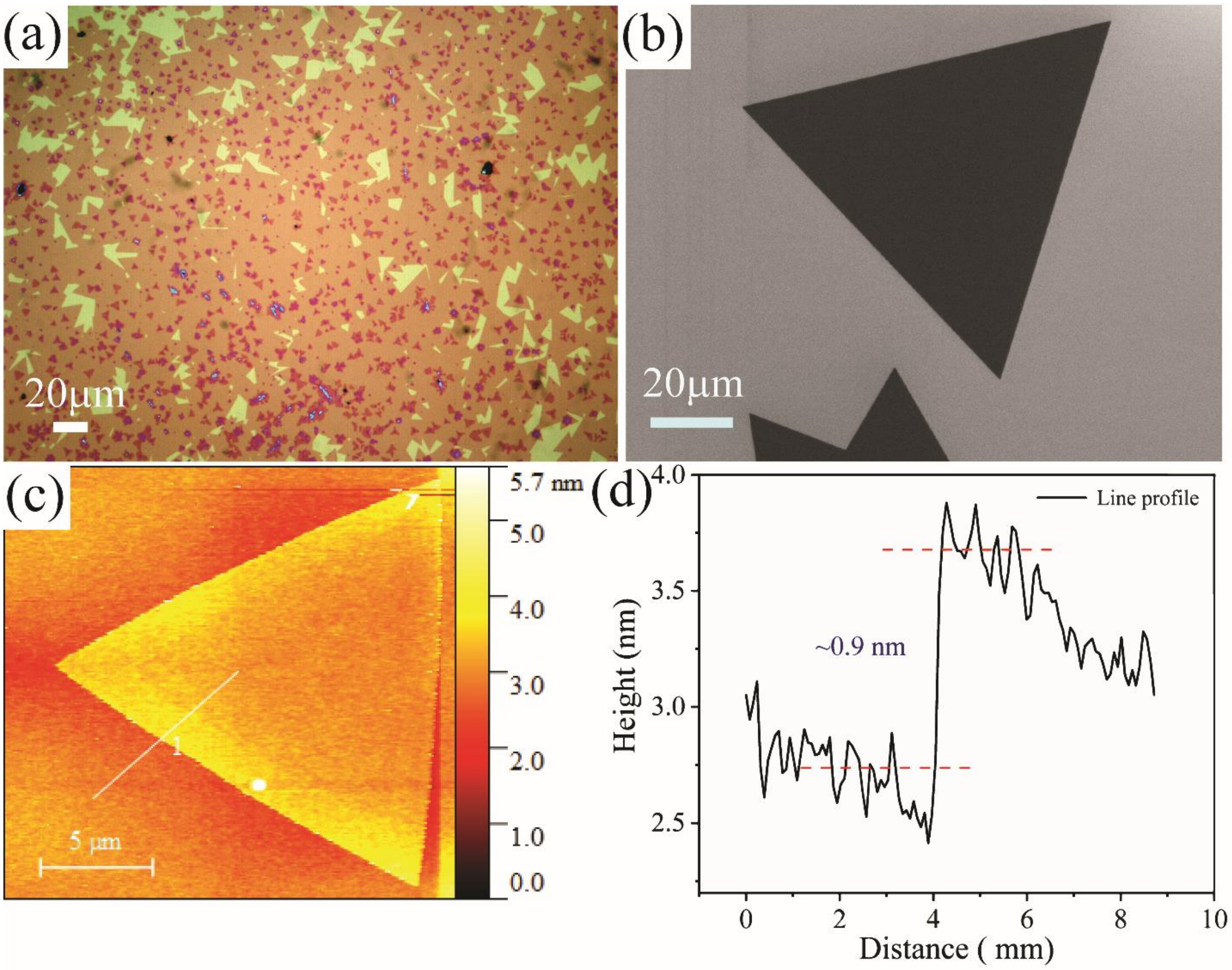

***Figure.1*** *(a) Shows optical micrograph of a grown continuous layer of* $WS_2$ *and (b) FESEM image of one of the isolated crystals of* $WS_2$ *(c) Atomic Force Microscope (AFM) image of single layer of* $WS_2$ *(d) Thickness profile of* $WS_2$ *isolated crystal showing height = 0.9 nm indicating monolayer growth.*

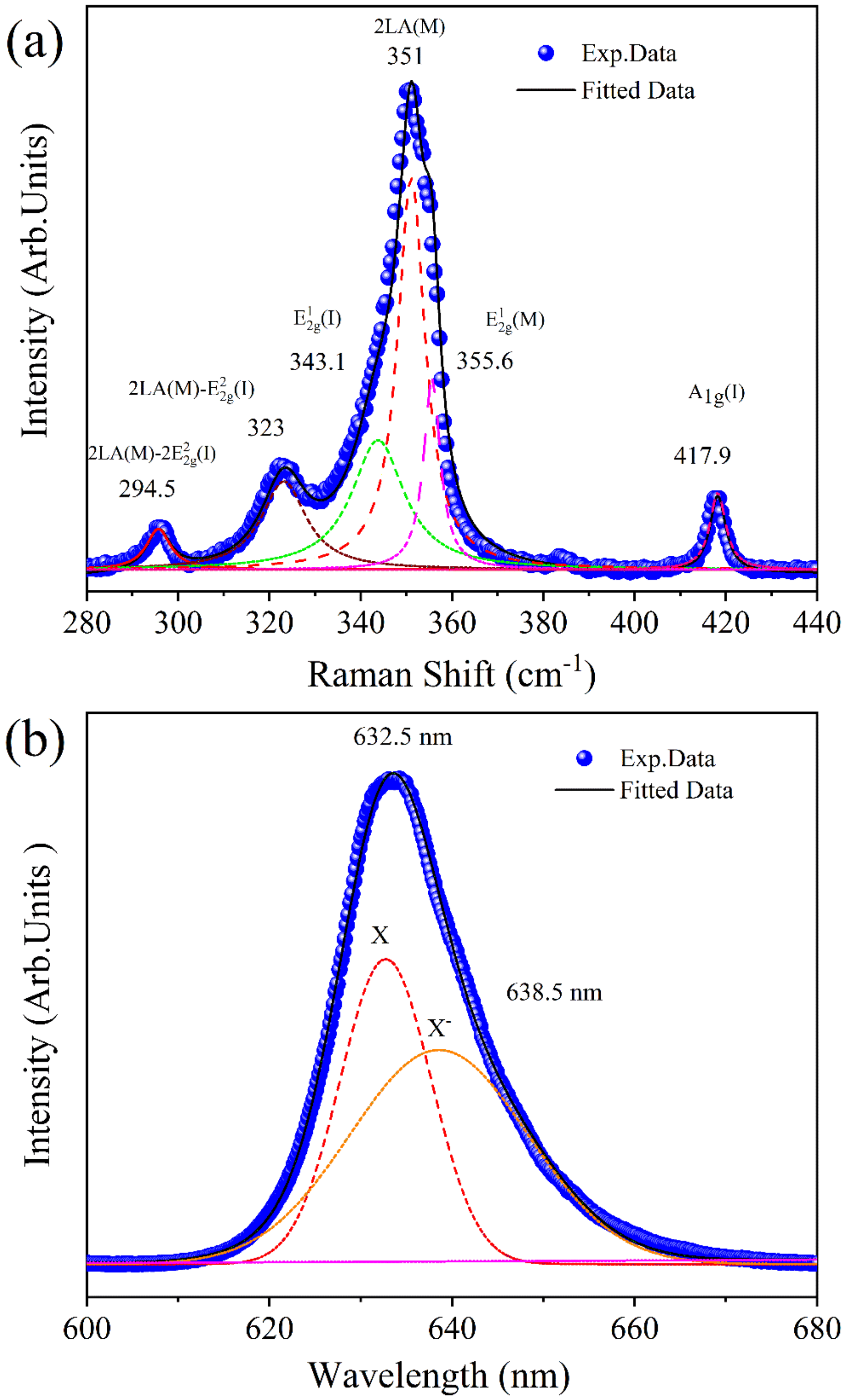

***Figure.2*** *(a) shows the Raman Spectra and (b) Photoluminescence spectra of the monolayer* $WS_2$ *grown sample.*

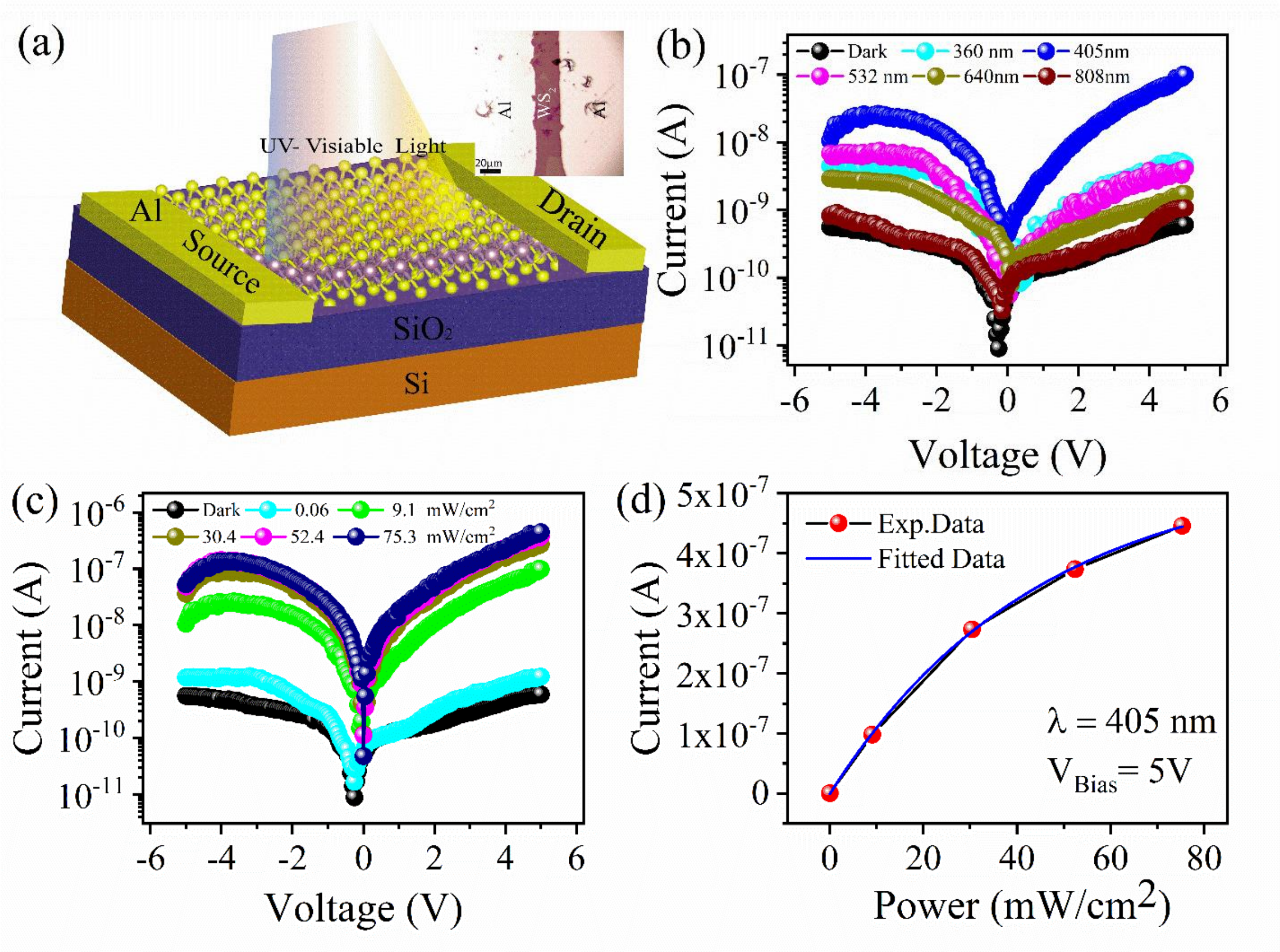

***Figure.3****(a) schematic diagram and optical Microscopy image of devices (b) I-V plot with different wavelength of laser source (c) I-V plot with varying incident intensity (under dark and upon 405 nm laser excitation (d) logarithmic plot of the photo-current as a function of laser power intensity and fitted with power law for photodetector at a bias voltage of 5V.*

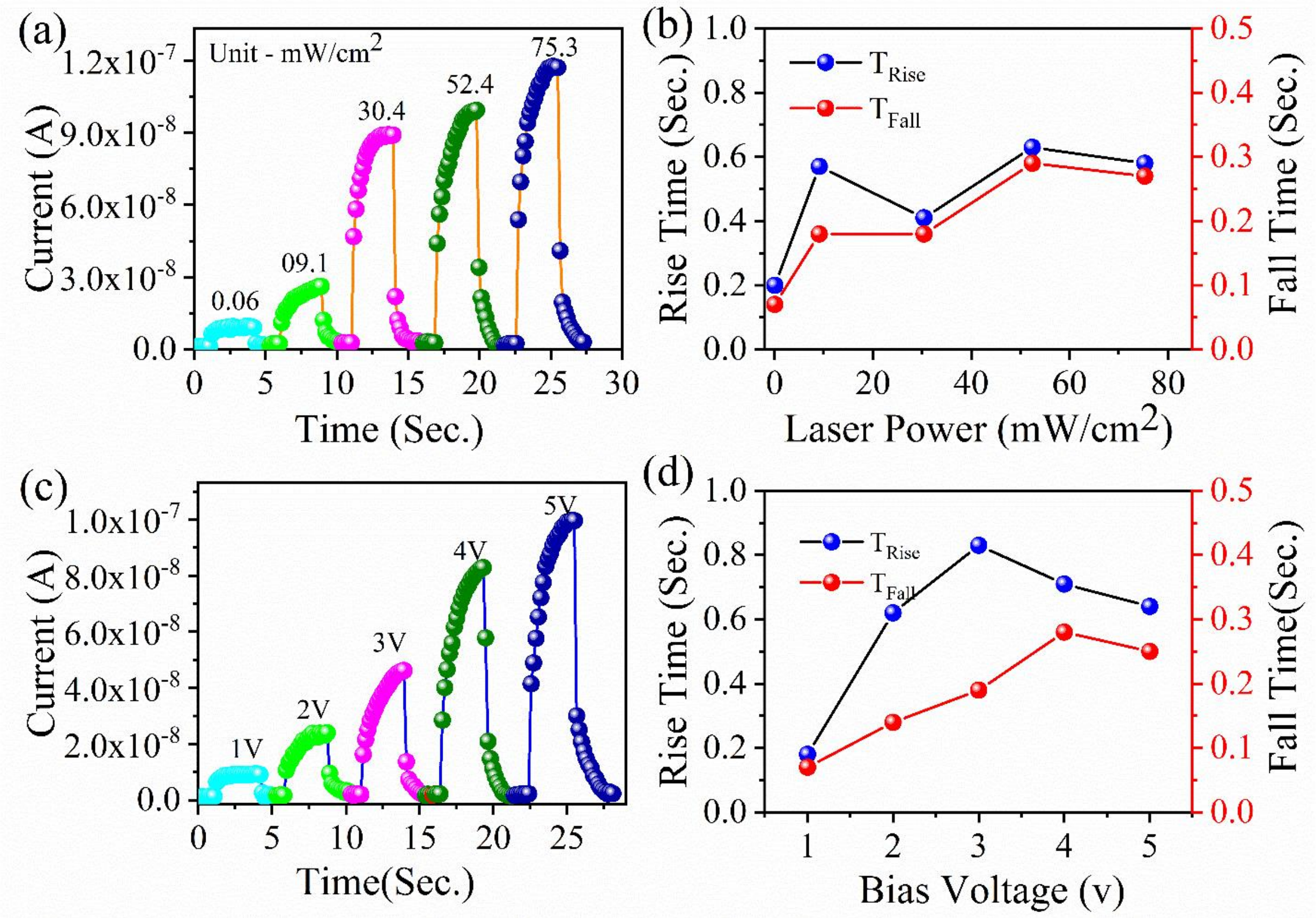

***Figure.4*** *(a) Photocurrent response with different illumination varying laser power density with fixed bias voltage 5V. (b) Change in Rise and fall time of photo current response with respect to the variation of pulse laser power density. (c) Voltage–dependent photocurrent response under fixed laser excitation of a 405 nm laser. (d) Change in Rise and fall time of Photo current response with respect to the variation in bias potential with constant illuminated laser power.*

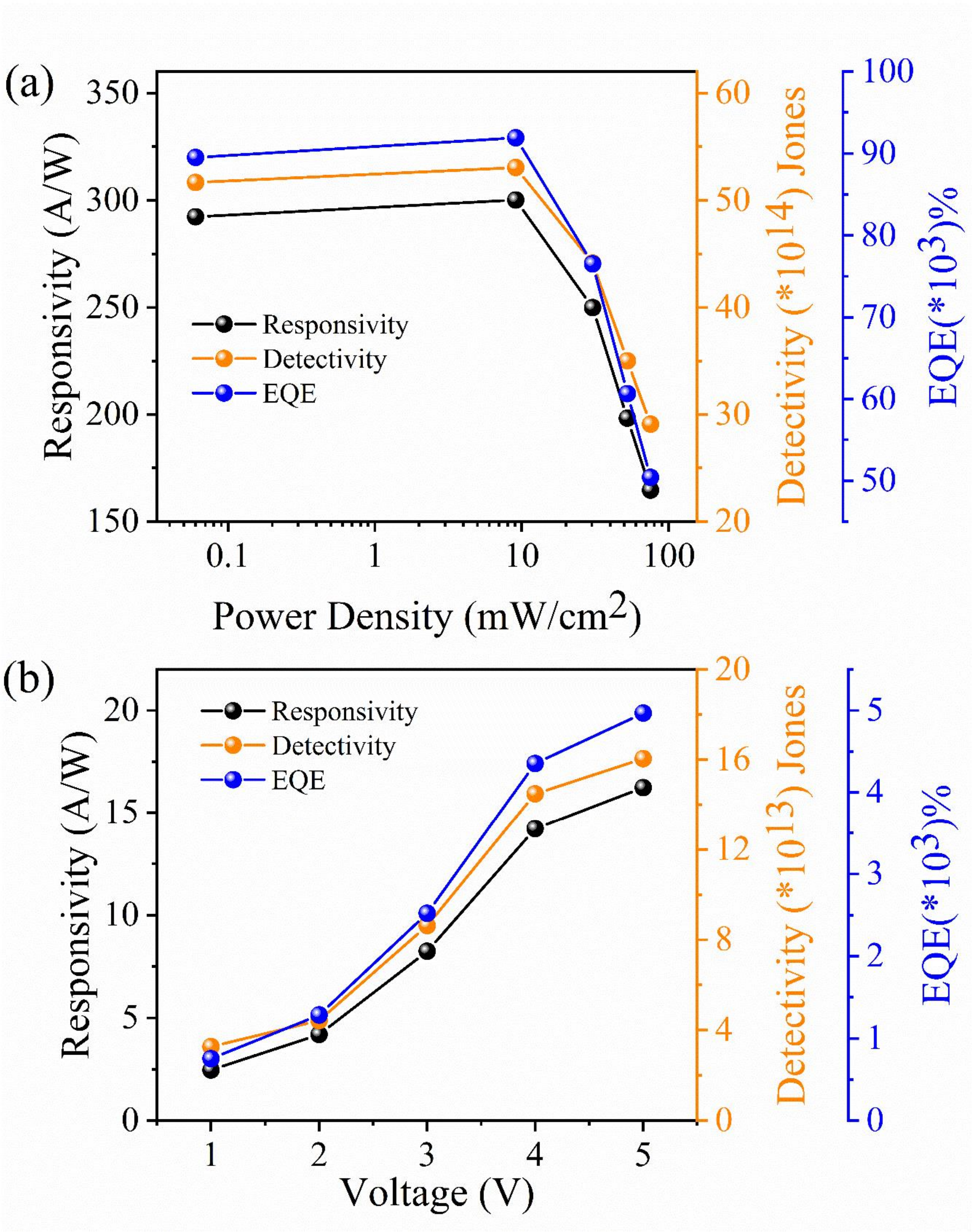

***Figure.5*** *Responsivity, detectivity and EQE of the device as a function of (a) illumination intensity at wavelength 405 nm and (b) bias voltage.*

## Supplementary Information

# High-Efficiency Photodetector Based On CVD-Grown $WS_2$ Monolayer

Rakesh K. Prasad[1], Koushik Ghosh[2], P. K. Giri[2], Dai-Sik Kim[3], Dilip K. Singh[1*]

[1]Department of Physics, Birla Institute of Technology Mesra, Ranchi -835215, India

[2]Department of Physics, Indian Institute of Technology Guwahati, Assam-781039, India

[3]Department of Physics and Quantum Photonics Institute and Center for Atom Scale Electromagnetism, Ulsan National Institute of Science and Technology (UNIST), Ulsan-44919, Republic of Korea

[*]Email: dilipsinghnano1@gmail.com

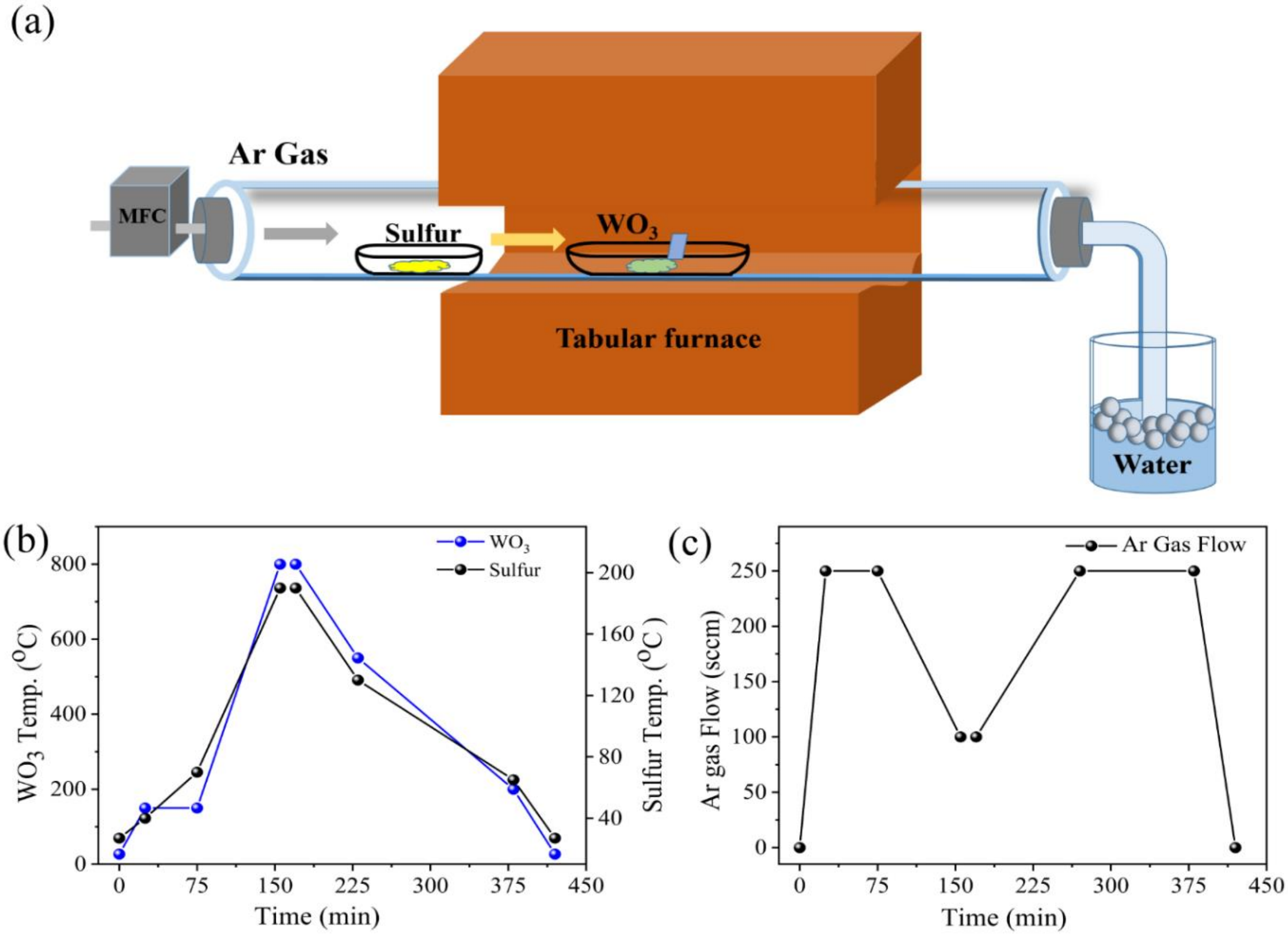

**Figure S1 (a)** Schematic diagram of Thermal CVD (b) Temperature and (c) Gas flow rate profile during the complete growth process.

Fig. S1 shows the schematic of the CVD setup used for growth. In CVD, the quartz tube of length 120 cm and diameter 4.5 cm was used. The temperature profile of both the precursors ($WO_3$ and S powder) during the complete cycle of growth has been shown in Fig. S1 (b) and the gas flow rate for the complete experimental duration. 285 nm $SiO_2$/Si wafer was used as a substrate for growth. The substrate was first cleaned sequentially using the isopropyl alcohol, water and ethanol through the sonication bath process for 15 min each. After sonication, the substrate was washed with deionized water and then dried using the argon gas and was placed inside the hot oven at 100 ℃ for 10 min. The polished surface of the substrate faces downward to the precursor above the $WO_3$ powder. Precursors and substrate were placed inside the CVD furnace as schematically shown in Fig. S1(a) separated at 19 cm. The outlet of the quartz tube was bubbled through a water bucket ensuring an equilibrium argon pressure inside the tube.

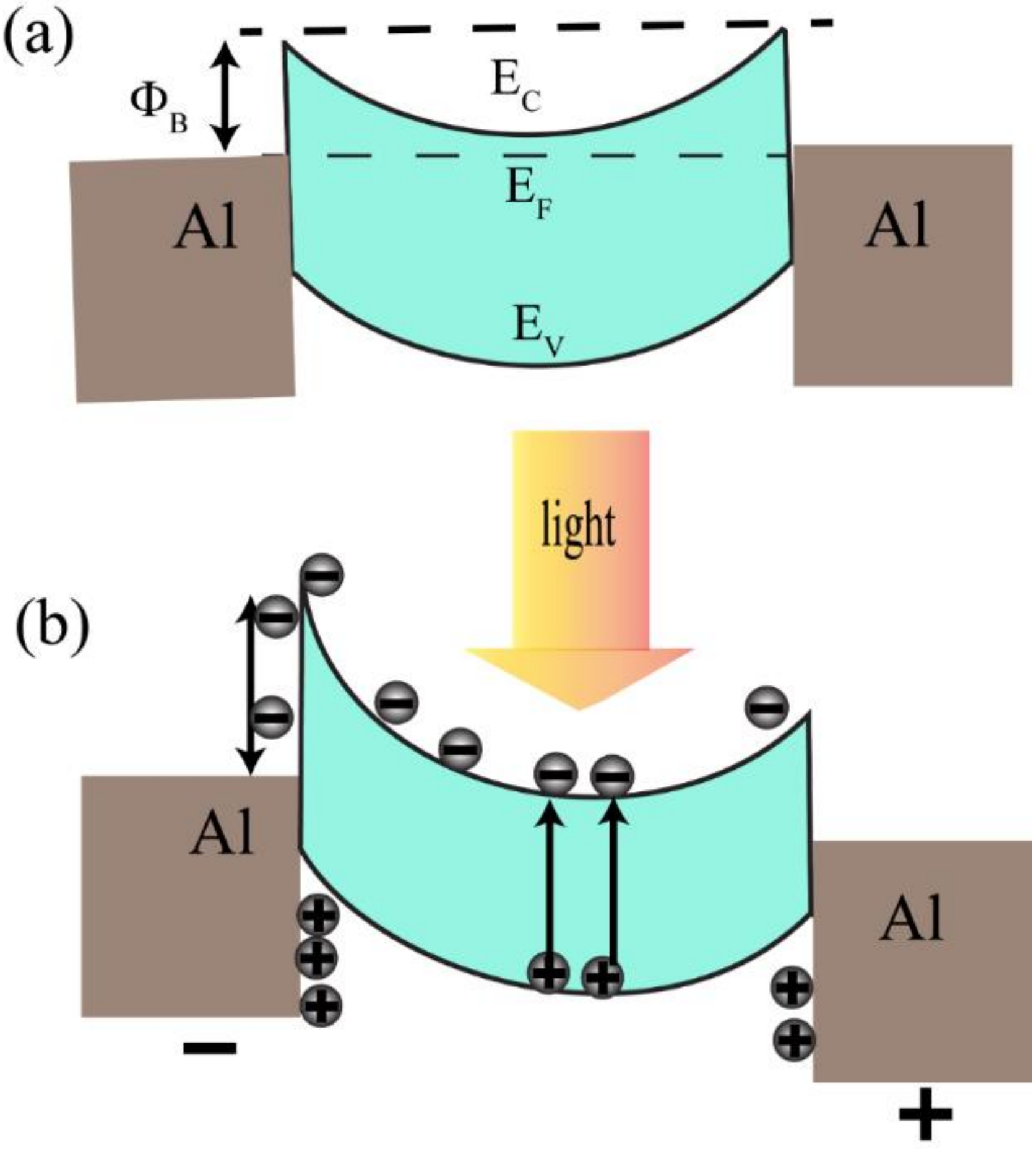

Fig. S2 Schematic diagram of mechanism of the photodetector (a) Without any bias and (b) with applied forward bias.

The band diagram for the metal-semiconductor-metal (M-S-M) contacts is schematically shown in Fig. S2

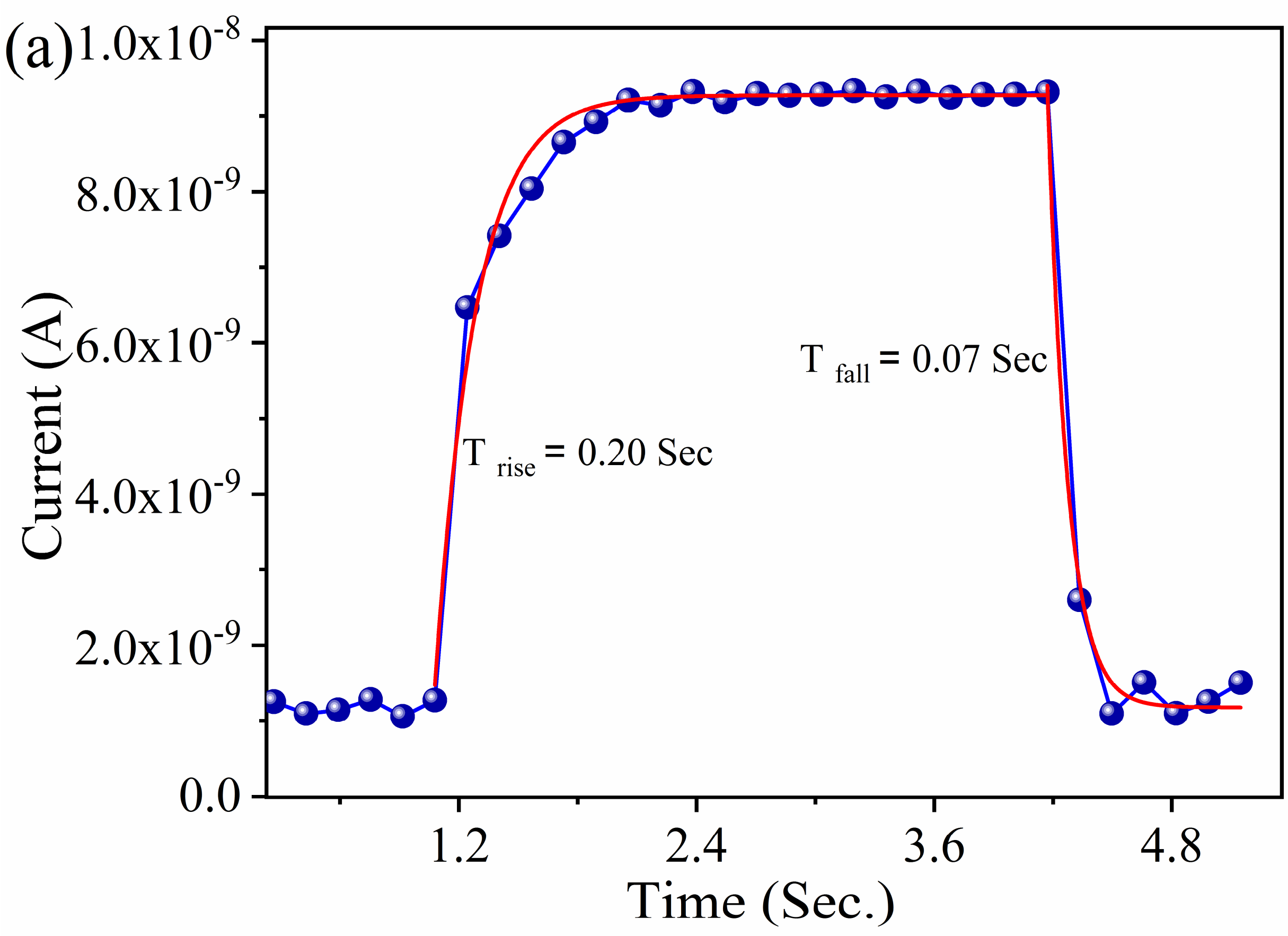

Figure S3: Rise Time (0.20 sec) and Fall Time (0.07 sec) of the device at 2Hz

**Table-SI** Comparison of the performance parameters of some common reference photodetectors based on Silicon, Ge and III-V nitrides and 2D-Semiconductors ($MoS_2$ and $WS_2$).

| *No. of Layer* | *Measurement Condition* | *Responsivity ($AW^{-1}$)* | *Detectivity ($D^*$) cm $Hz^{1/2}$ $W^{-1}$* | *Response Time (µs)* | *Reference* |
|---|---|---|---|---|---|
| Si | λ = 254nm | 0.14 | N/A | 0.2-5.9 | [1] |
| Si | λ = 850nm-900nm | N/A | ~$4\times10^{13}$ | N/A | [2] |
| Multilayer $WS_2$ | λ = 514nm | $92\times10^{-4}$ | N/A | $53\times10^{2}$ | [3] |
| Multilayer $WS_2$ (PLD grown) | - | 0.51 A $W^{-1}$ | $2.7\times10^{9}$ | - | [4] |
| 2L-$WS_2$ | $V_{ds}$= 1V<br>$V_g$ = 0V<br>λ = 457nm<br>P= 0.5 $mWcm^{-2}$ | 3106 | $5\times10^{12}$ | 59(rise)<br>87(decay) | [5] |
| 1l-$WS_2$ | $V_{ds}$= 10V<br>$V_g$ = 0V<br>λ = 532nm<br>P= 0.07 $mWcm^{-2}$ | $5.2\times10^{-4}$ | $4.9\times10^{9}$ | < 560(rise)<br><560 (decay) | [6] |
| 1L -$WS_2$ | $V_{ds}$= 20V<br>$V_g$ = 60V<br>λ = 532nm<br>P= 70 mW | 0.019 | N/A | $6\times10^{4}$<br>$1.9\times10^{5}$ | [7] |
| 1L-$WS_2$ | λ = 500nm | 3.07 | N/A | $37\times10^{4}$ | [8] |
| 1L – $WS_2$ | λ = 405nm | 290 | $52*10^{14}$ | $2.0\times10^{4}$<br>$7.0\times10^{3}$ | Our work |